\documentclass[aps,preprint,onecolumn,showpacs,floatfix,superscriptaddress,noshowpacs,longbibliography]{revtex4-1}
\usepackage{graphicx,graphics}
\usepackage{amsmath,amssymb,amsfonts}
\usepackage{color}
\usepackage[colorlinks=true,citecolor=blue,linkcolor=blue,urlcolor=blue]{hyperref}

\usepackage{xspace}
\usepackage{tabularx}

\begin{document}

\title{Large magnetoresistance in the iron-free pnictide superconductor LaRu$_2$P$_2$}

\author{Marta Fern\'andez-Lomana}
\affiliation{Laboratorio de Bajas Temperaturas y Altos Campos Magn\'eticos, Departamento de F\'isica de la Materia Condensada, Instituto Nicol\'as Cabrera and Condensed Matter Physics Center (IFIMAC), Unidad Asociada UAM-CSIC, Universidad Aut\'onoma de Madrid, E-28049 Madrid,
Spain}

\author{V\'ictor Barrena}
\affiliation{Laboratorio de Bajas Temperaturas y Altos Campos Magn\'eticos, Departamento de F\'isica de la Materia Condensada, Instituto Nicol\'as Cabrera and Condensed Matter Physics Center (IFIMAC), Unidad Asociada UAM-CSIC, Universidad Aut\'onoma de Madrid, E-28049 Madrid,
Spain}

\author{Beilun Wu}
\affiliation{Laboratorio de Bajas Temperaturas y Altos Campos Magn\'eticos, Departamento de F\'isica de la Materia Condensada, Instituto Nicol\'as Cabrera and Condensed Matter Physics Center (IFIMAC), Unidad Asociada UAM-CSIC, Universidad Aut\'onoma de Madrid, E-28049 Madrid,
Spain}

\author{Sara Delgado}
\affiliation{Laboratorio de Bajas Temperaturas y Altos Campos Magn\'eticos, Departamento de F\'isica de la Materia Condensada, Instituto Nicol\'as Cabrera and Condensed Matter Physics Center (IFIMAC), Unidad Asociada UAM-CSIC, Universidad Aut\'onoma de Madrid, E-28049 Madrid,
Spain}

\author{Federico Mompe\'an}
\affiliation{Instituto de Ciencia de Materiales de Madrid, Consejo Superior de
Investigaciones Cient\'{\i}ficas (ICMM-CSIC), Unidad Asociada UAM-CISC, Sor Juana In\'es de la Cruz 3,
28049 Madrid, Spain}

\author{Mar Garc{\'i}a-Hern{\'a}ndez}
\affiliation{Instituto de Ciencia de Materiales de Madrid, Consejo Superior de
Investigaciones Cient\'{\i}ficas (ICMM-CSIC), Unidad Asociada UAM-CISC, Sor Juana In\'es de la Cruz 3,
28049 Madrid, Spain}

\author{Hermann Suderow}
\affiliation{Laboratorio de Bajas Temperaturas y Altos Campos Magn\'eticos, Departamento de F\'isica de la Materia Condensada, Instituto Nicol\'as Cabrera and Condensed Matter Physics Center (IFIMAC), Unidad Asociada UAM-CSIC, Universidad Aut\'onoma de Madrid, E-28049 Madrid,
Spain}

\author{Isabel Guillam\'on}
\affiliation{Laboratorio de Bajas Temperaturas y Altos Campos Magn\'eticos, Departamento de F\'isica de la Materia Condensada, Instituto Nicol\'as Cabrera and Condensed Matter Physics Center (IFIMAC), Unidad Asociada UAM-CSIC, Universidad Aut\'onoma de Madrid, E-28049 Madrid,
Spain}

\begin{abstract}
The magnetoresistance of iron pnictide superconductors is often dominated by electron-electron correlations and deviates from the H$^2$ or saturating behaviors expected for uncorrelated metals. Contrary to similar Fe-based pnictide systems, the superconductor LaRu$_2$P$_2$ (T$_c$ = 4 K) shows no enhancement of electron-electron correlations. Here we report a non-saturating magnetoresistance deviating from the H$^2$ or saturating behaviors in LaRu$_2$P$_2$. We have grown and characterized high quality single crystals of LaRu$_2$P$_2$ and measured a magnetoresistance following H$^{1.3}$ up to 22 T. We discuss our result by comparing the bandstructure of LaRu$_2$P$_2$ with Fe based pnictide superconductors.  The different orbital structures of Fe and Ru leads to a 3D Fermi surface with negligible bandwidth renormalization in  LaRu$_2$P$_2$, that contains a large open sheet over the whole Brillouin zone. We show that the large magnetoresistance in LaRu$_2$P$_2$ is unrelated to the one obtained in materials with strong electron-electron correlations and that it is compatible instead with conduction due to open orbits on the rather complex Fermi surface structure of LaRu$_2$P$_2$.
\end{abstract}

\maketitle

\section*{Introduction}

Magnetoresistance (MR) in unconventional high critical temperature superconductors such as cuprate and pnictide materials has been recently used as a measure of their unconventional electronic properties\cite{hayes2016scaling,hayes2018magnetoresistance,giraldo2018scale,ping2009magnetoresistance}. MR appears due to the changes in the electronic transport induced by the magnetic field (H), it is insensitive to a change in the direction of the applied magnetic field and therefore increases with an even power in H, which turns out to be H$^2$ in most cases. Deviation from the H$^2$ behaviour towards linear magnetoresistance in optimally doped cuprates and pnictide superconductors has been interpreted as a signature of the strange metal state associated to the high T$_c$ in these materials\cite{hayes2016scaling,hayes2018magnetoresistance,giraldo2018scale}. The Fermi surface of the cuprates is particularly simple, with pockets at the corners of the Brillouin zone that evolve into arcs with doping into the superconducting state\cite{marshall1996unconventional}. In the pnictides, there are many bands crossing the Fermi level. However, in most cases the bands can be grouped into hole pockets centered at the Brillouin zone and electron pockets at the corner, the latter often being quasi-two dimensional. In stoichiometric non-superconducting pnictides, positive large MR proportional to H$^n$ with n$<$1.5 has been reported as a consequence of magnetic field enhanced spin-density-wave gap\cite{ping2009magnetoresistance}. Here we present MR experiments up to 22 T in the stoichiometric pnictide superconductor LaRu$_2$P$_2$.  We show that LaRu$_2$P$_2$ has a positive non-saturating MR with a field dependence close to linear and similar in magnitude to the one obtained in Fe-based superconductors. However its origin is not due to electronic correlations nor the presence of a spin-density wave, both absent in LaRu$_2$P$_2$, but is instead originated by he contribution from open orbits to the MR. We adscribe this to the presence of an additional large open Fermi surface sheet in LaRu$_2$P$_2$ induced by the different orbital structures of Fe and Ru, also responsible for the low critical temperature and lack of electronic correlations which makes the difference between LaRu$_2$P$_2$ and Fe-based systems.

LaRu$_2$P$_2$ crystallizes in the ThCr$_2$Si$_2$-type structure (space group I4/mmm). It is isostructural with the 122 Fe pnictides XFe$_2$As$_2$ where X = Ba, Ca or Sr. Neither magnetic nor structural transition have been reported in this compound at ambient pressure. Superconducting properties are isotropic and the critical temperature  estimated from calculations of electron-phonon coupling is in good agreement with the experimental value (T$_c$ = 4.1 K), suggesting electron-phonon mediated superconductivity\cite{jeitschko1987superconducting,ying2010isotropic,karaca2016first}. Superconducting and structural properties are modified by the effect of hydrostatic pressure. T$_c$ first increases to above 5 K and then it is suddenly suppressed at 2 GPa\cite{foroozani2014hydrostatic,li2016pressure}. At similar hydrostatic pressures, LaRu$_2$P$_2$ undergoes a collapsed tetragonal phase transition suggesting that disappearance of superconductivity is likely driven by the structural transition to the collapsed tetragonal phase\cite{drachuck2017collapsed}. 

The Fermi surface and bandwidth renormalization of LaRu$_2$P$_2$ are very different from their counterparts in the superconducting Fe pnictides. The substitution of Fe-3d by less localized Ru-4d orbitals increases the bandwidth producing larger Fermi surfaces with reduced electronic correlations \cite{brouet2010significant}. In contrast to quasi-two-dimensional electronic properties found in most pnictide superconductors, DFT calculations obtained a three dimensional Fermi surface for LaRu$_2$P$_2$ \cite{moll2011quantum,razzoli2012bulk}.  It consists of two sets of sheets derived from the Ru-4d orbitals that form a donut with a small hole centered around the M point and a warped electron cylinder at the corners of the Brillouin zone and an additional open three dimensional sheet derived from a combination of La and Ru-4d orbitals\cite{moll2011quantum,razzoli2012bulk}. The electronic structure determined experimentally by de Haas-van Alphen oscillations and angle-resolved photoemission agrees well with DFT calculations showing that the bandwidth renormalization is negligible\cite{moll2011quantum,razzoli2012bulk}. They find mass enhancement close to 1, similar to other non-superconducting pnictides as LaFe$_2$P$_2$, indicating that superconductivity in LaRu$_2$P$_2$ is conventional in the sense that it is not attributed to electron-electron interaction\cite{blackburn2014fermi}. 

Here we report single crystal growth and characterization of the superconducting properties of LaRu$_2$P$_2$ and study the MR up to 22 T. We find large and non-saturating MR up to 22 T and show that this can be explained by the contribution from open orbits to the MR. We compare our results in LaRu$_2$P$_2$ with those obtained in superconducting Fe pnictides and discuss the different the origin of the MR in both systems.

\section*{Experimental}

Single crystals of LaRu$_2$P$_2$ were obtained using the solution growth method\cite{Canfieldbook,Canfield01,CanfieldFisk,CanfieldNewMat}. We used La (Alfa Aesar 99.9 $\%$), Ru (Goodfellow 99.9 $\%$), red P (Alfa Aesar 99.999 $\%$ metal basis) and added Sn (Goodfellow 99.995 $\%$) as flux in a total molar ratio of 1:2:2:15 (La:Ru:P:Sn). We mixed all the elements into a standard CCS alumina crucible configuration\cite{canfield2016use}. The crucibles were sealed into a silica ampoule filled with a residual Ar atmosphere below 10$^{-3}$ mbar. The ampoule was heated to 320$^o$C  and kept at this temperature for 3 hours to allow for inclusion of P into liquid Sn (see Reference \cite{drachuck2017collapsed}) and avoid overpressures due to the vapour pressure increase of P with temperature. We then continued the temperature ramp up, going up to 1195$^o$C in 3 hours. After 24 hours at that temperature, we cooled slowly to 750$^o$C in 288 hours. Then, we rapidly extracted the ampoule from the furnace and decanted the Sn flux with a centrifuge. Crystals were easily identified with clear geometric shapes that stood out in the growth pellet. Platelet-like single crystals of LaRu$_2$P$_2$ with typical size 0.5 $\times$ 0.5 $\times $0.1 mm$^3$ (top left inset in figure  \ref{Fig2}) were removed and carefully prepared by cleaving and shaping to eliminate all the residual Sn flux that was optically observed as small superficial droplets.

Upper left inset in Figure \ref{Fig2} shows X-ray diffraction measurements (XRD) in a cleaved single crystals of LaRu$_2$P$_2$ using Cu K$^{\alpha}$ radiation. The diffraction pattern shows a peak corresponding to the crystal plane (004) of LaRu$_2$P$_2$ at a characteristic angle of 33.6$^o$. Using Bragg's law, we find a value for the lattice parameter along the c-axis of 1.066 nm, in agreement with previous XRD measurements\cite{jeitschko1987superconducting}. We could only observe the (004) diffraction peak due to the reduced size of the single crystals and the crystallographic c-axis being perpendicular to the plane of the single crystals. No trace of Sn is observed. The cleaved crystals were then cut into a bar shape (0.1 $\times$ 0.4 $\times$ 0.05 mm$^3$) for transport measurements. Resistivity measurements have been performed using the four probe AC method,  with four electrical contacts made by gluing 25 $\mu$m gold wires with silver epoxy on the cleaved surface of LaRu$_2$P$_2$ samples.  For magnetoresistance measurements, we have used a cryostat with base temperature 1 K and equipped with a 20+2 T superconducting magnet\cite{22tmagnet,wu2020huge,montoya2019methods}.  Resistivity was measured using a SR 830 lock-in amplifier. The reference voltage of the lock-in was converted to a current using a Howland pump scheme\cite{CurrentSource}. Magnetization measurements have been made using a Quantum Design Squid magnetometer varying the magnetic field from 0 T to 0.1 T in samples with geometry 0.5 $\times$ 0.5 $\times$ 0.05  mm$^3$. All results shown here have been obtained by applying the magnetic field parallel to the c axis and perpendicular to the current. 

%

\section{Results}

 \begin{figure}[t]
 \begin{center}
\includegraphics[width=0.7\textwidth]{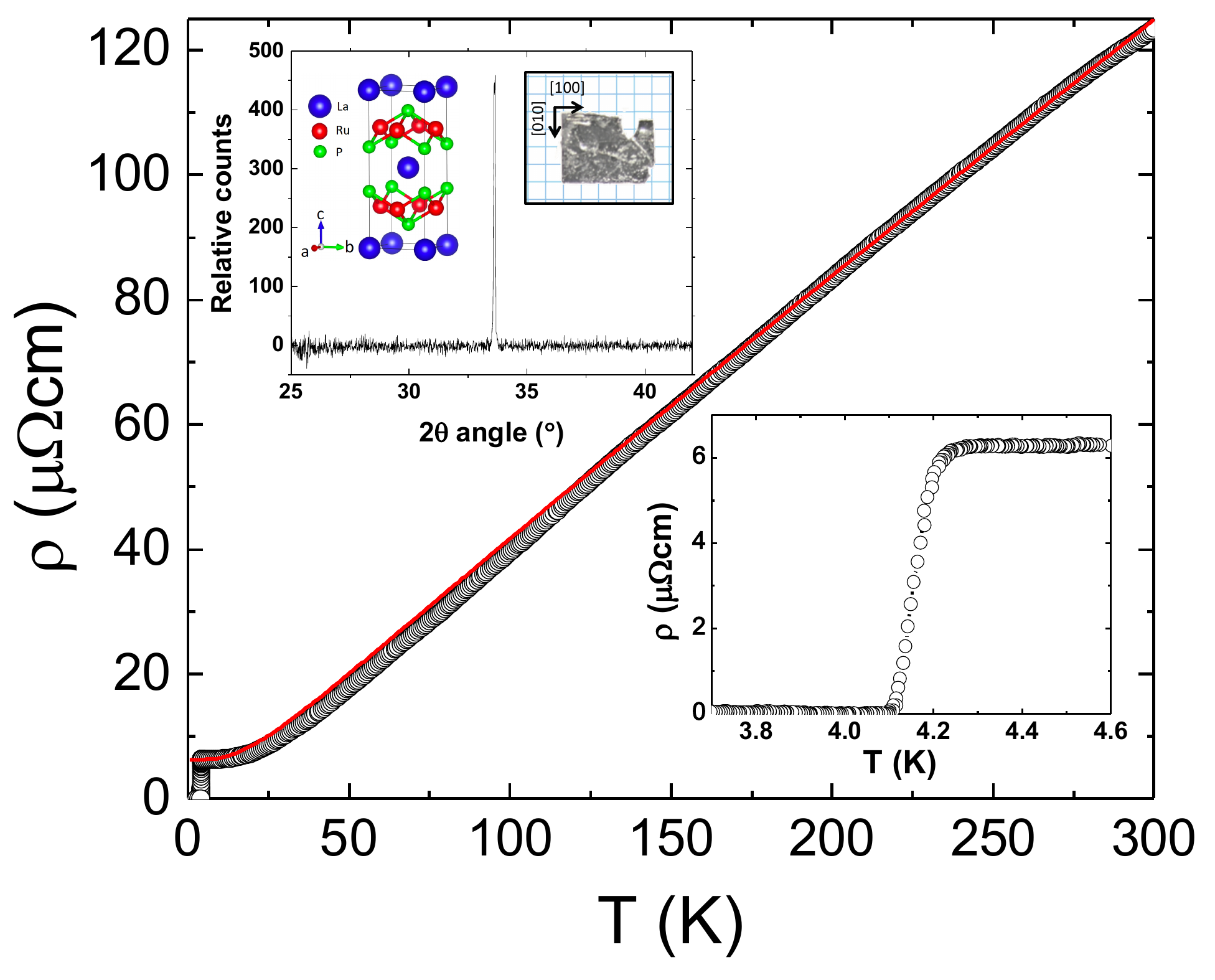}

\caption{In the main panel we show the resistivity of LaRu$_2$P$_2$ vs temperature as black points. Red line is the fit to Bloch-Gr\"uneisen expression given by Eq.\ref{Eq0}. In the bottom right inset we show a zoom up into the low temperature region, with the superconducting transition. In the upper left we show the X-Ray diffraction pattern of a LaRu$_2$P$_2$ single crystal using Cu K$^{\alpha}$ radiation (black line). We clearly observe the (004) peak  at a characteristic angle of 33.6$^o$. The inset also shows a photograph of one of the obtained LaRu$_2$P$_2$ single crystals with arrows indicating the [100] and [010] directions and the unit cell of LaRu$_2$P$_2$, which adopts the tetragonal ThCr$_2$Si$_2$-type structure (I4/mmm space group).
\label{Fig2}}
\end{center}
\end{figure}

Figure \ref{Fig2} shows the temperature dependence of the resistivity $\rho$ in LaRu$_2$P$_2$. $\rho(T)$ drops with decreasing temperature as expected for a good metal and saturates below 10 K with a zero temperature extrapolation $\rho_0$ = 6.3 $\mu \Omega$cm. The residual resistivity ratio of our samples, $RRR$ = $\rho_{\text{300K}}/\rho_{\text{10K}}$, is above 20, similar as obtained previously in high quality single crystalline samples of LaRu$_2$P$_2$\cite{ying2010isotropic,moll2011quantum,li2016pressure} and evidences the absence of large amounts of defects or inclusions. We estimate the electronic mean free path to be above 20 nm. We use the Eliashberg spectral function $\alpha^2$F($\omega$) from Ref.\cite{karaca2016first} to calculate the temperature dependence of the resistivity in LaRu$_2$P$_2$ via the Bloch-Gr\"uneisen (BG)  expression
\begin{equation}
\rho_{BG} = \rho_0 + \frac{4\pi m}{n e^2} \int^{\omega_{max}}_0 \alpha^2F(\omega) \frac{xe^x}{(e^x-1)^2} d\omega \label{Eq0},
\end{equation}
where we use $\rho_0$ = 6.3 $\mu \Omega$cm, m is the free electron mass, n= 3$\times$10$^{22}$ cm$^{-3}$ is the carrier density obtained by Hall effect measurements in LaRu$_2$P$_2$\cite{ying2010isotropic}, e is the electron charge and x = $\omega$/T with $\omega$ being the phonon frequency. We take $\hbar\omega_{max}$ = k$_B$T$_D$, with the Debye temperature $\Theta_D$ = 392 K obtained by density functional theory calculations\cite{rahaman2017thcr2si2}. The calculated curve is shown by the red line in figure \ref{Fig2}. The expression \ref{Eq0} provides an excellent account of the whole temperature dependence of the resistivity. At high temperatures, where $\rho$ is linear with T, the electron-phonon coupling constant $\lambda$ is related with the resistivity slope by $\lambda$ = ($\hbar\omega_p^2/8\pi^2$k$_B$)(d$\rho$/dT), with $\omega_p$ the plasma frequency\cite{allen1978new}. We obtained  $\lambda \approx$ 1 which is similar to the value calculated by integrating the Eliashberg spectral function used to obtain the red curve in figure \ref{Fig2}\cite{karaca2016first} and the value obtained experimentally from quantum oscillations\cite{moll2011quantum}. At low temperatures, we observe superconductivity with  T$_c$ = 4.16 K (upper inset in figure \ref{Fig2}), similar to the T$_c$ obtained in other high quality single crystalline samples\cite{jeitschko1987superconducting,ying2010isotropic}. As mentioned above, the critical temperature estimated from $\lambda$ is in good agreement with the experiments, suggesting electron-phonon mediated superconductivity\cite{karaca2016first}.

\begin{figure}[t]
\begin{center}
\includegraphics[width=1\textwidth]{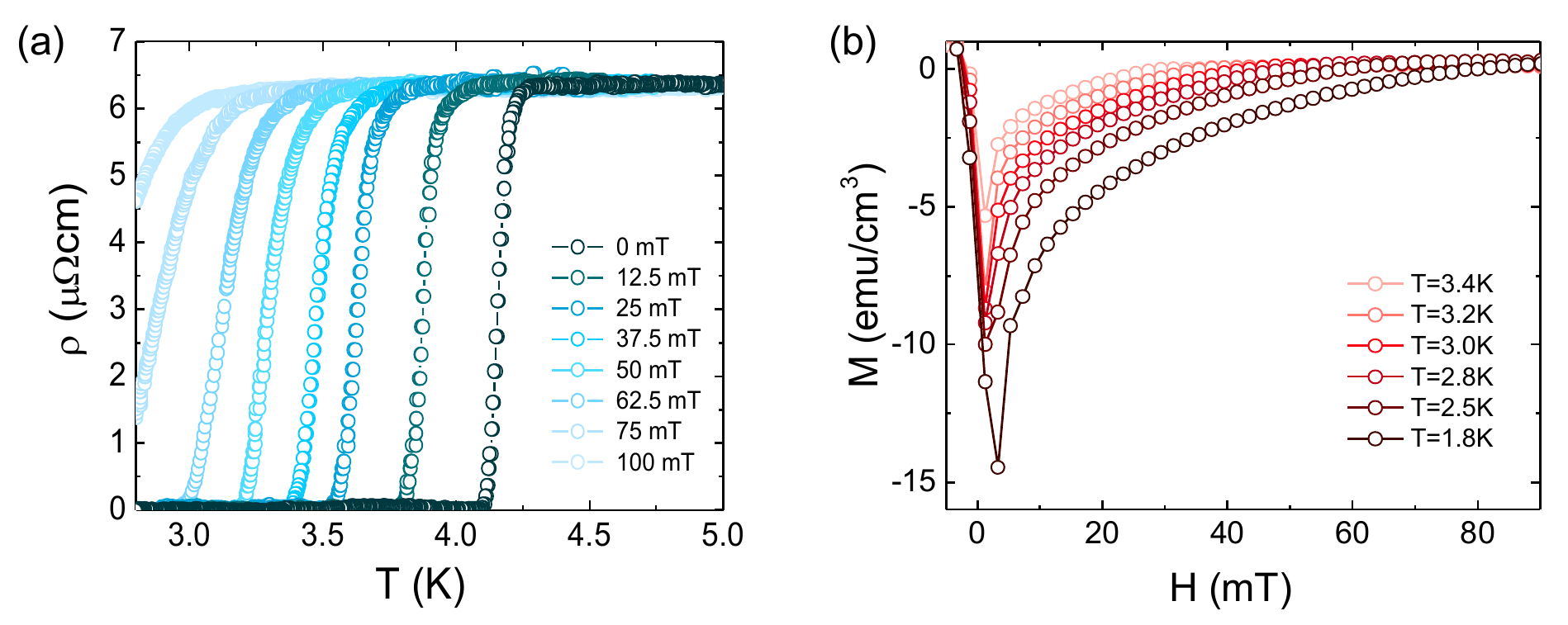}
		
\caption{(a) We show as points the resistivity vs temperature from zero to 0.1 T. Colors indicate the value of the magnetic field, provided in the legend. (b) We show as points the magnetization as a function of the magnetic field. Colors indicate the measurement temperatures, given in the legend.\label{Fig3}}
\end{center}
\end{figure}

\begin{figure}[t]
\begin{center}
\includegraphics[width=0.7\textwidth]{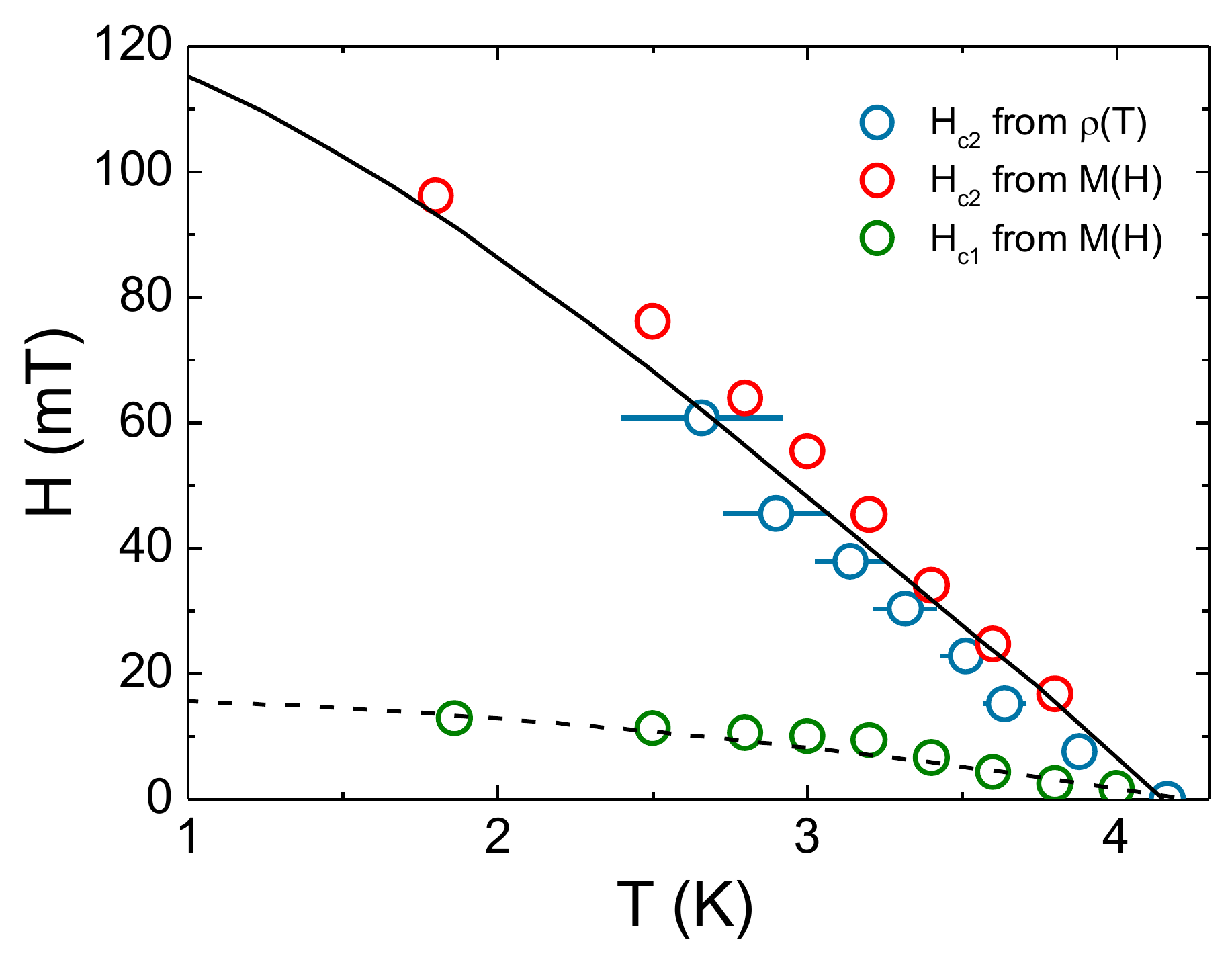}		
\caption{Lower and upper critical fields, H$_{c1}$ and H$_{c2}$, as a function of temperature. Blue and red points are experimental data of H$_{c2}$ obtained from the superconducting transition in resistivity and magnetization measurements shown in figure \ref{Fig3}. Blue points correspond to the temperature at which $\rho$ is $50\%$ of $\rho_0$ and the error bars represent the temperature range from 10$\%$ to 90$\%$$\rho_0$ (at the highest temperatures the error is smaller than the point size). The continuous black line is the fit to the WHH theory. Green points are H$_{c1}$ obtained from magnetization experimentes as described in the text. Black dotted line is the fit to H$_{c1}$(T) = H$_{c1}$(0) [1 - (T/T$_c$)$^2$] with H$_{c1}$(0) = 0.017T.  \label{Fig4}}
\end{center}
\end{figure}

In figure \ref{Fig3}(a,b) we show the temperature dependence of the resistivity at different magnetic fields and the field dependence of the magnetization at different temperatures . We can extract the temperature dependence of the upper critical field H$_{c2}$ from the resistive transition and from the magnetization. We show the result in Figure  \ref{Fig4} (blue and red open points). H$_{c2}$ follows well the Werthammer-Helfand-Hohenberg (WHH) formula (black line)\cite{Werthamer66}. With the zero temperature extrapolation H$_{c2}$(0) = 0.127 T we obtain $\xi$ = 50 nm, using H$_{c2}$(0) = $\phi_0$/2$\pi\mu_0\xi^2$ where $\phi_0$ = 2.07 $\times$ 10$^{-15}$ Wb is the flux quantum. This value is quite large, as compared with Fe based pnictide superconductors, which show coherence lengths that are usually of a few nm. We can also extract the temperature dependence of H$_{c1}$ (green open points in \ref{Fig4}). To this end, we determine the field for entry of vortices H$_0$  and calculate H$_{c1}$ taking into account the demagnetizing factor n a rectangular sample (H$_{c1}$ = H$_0$/tanh$\sqrt{0.36 d/a}$, where $d$ is the sample dimension along the field and $a$ perpendicular to the field\cite{brandt1999irreversible}) at each temperature. We fit the temperature dependence to the usual phenomenological expression H$_{c1}$(T) = H$_{c1}$(0)[1-(T/T$_c$)$^2$] (the black dotted line in \ref{Fig4}) and obtain the zero temperature extrapolation H$_{c1}$(0) = 0.017 T. We can then estimate the penetration depth using H$_{c1}$(0) = ($\phi_0$/4$\pi\mu_0\lambda^2$)(ln$\kappa$+0.5) where $\kappa$ = $\lambda/\xi$ is the Ginzburg-Landau parameter. We find $\lambda$ = 83.4 nm and $\kappa$ = 1.5, showing that LaRu$_2$P$_2$ is a weakly type-II superconductor.

\begin{figure}[t]
\begin{center}
\includegraphics[width=0.7\textwidth]{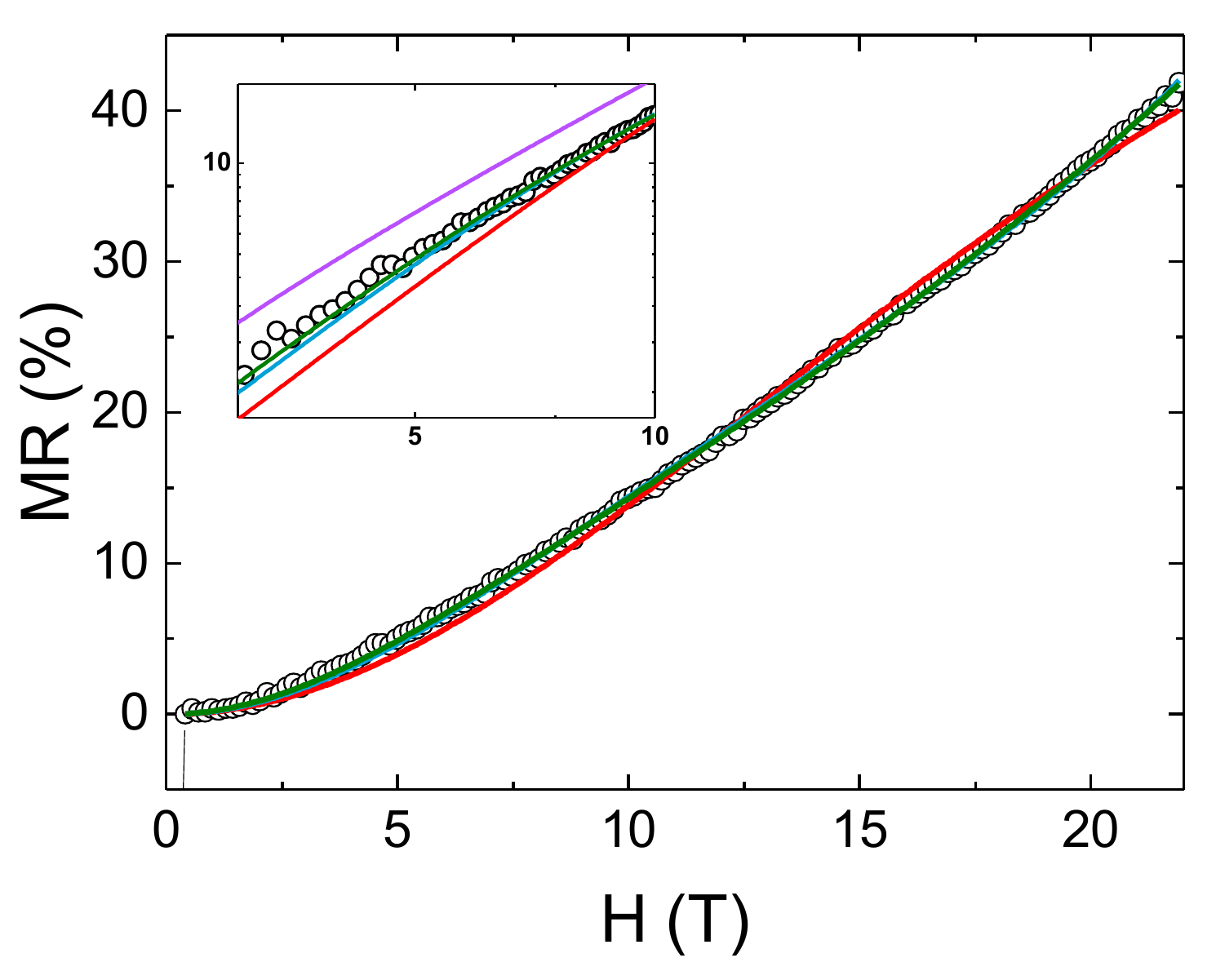}
\caption{Magnetoresistance defined as MR = ($\rho$(H)/$\rho_0$ -1) $\times$ 100\% in LaRu$_2$P$_2$ up to 22 T (black points). Red line is the fit to two-band galvanomagnetic effect model (Eq.\ref{Eq2}) with $\alpha$ = 6.9$\pm$0.1 and $\mu$ = 0.0615$\pm$0.0006 m$^2$V$^{-1}$s$^{-1}$. Blue and green lines are the fits to the same model but including, respectively, the magnetic field effect on CDW or SDW (Eq.\ref{Eq3}) with $\alpha$ = 17.8$\pm$0.2, $\beta$ = 0.39$\pm$0.01 and $\mu$ = 0.1$\pm$0.001 m$^2$V$^{-1}$s$^{-1}$, and the contribution from open orbits (Eq.\ref{Eq4}) with $\alpha$ = 23.3$\pm$0.5, $\mu$ = 0.110$\pm$0.001 m$^2$V$^{-1}$s$^{-1}$ and $\delta$ = (425$\pm$7)10$^{-4}$. The inset shows MR below 10 T in logarithmic scale. Red, blue and green lines are the fits shown in the main panel and purple line is the proportional to H$^n$ with n = 1.32.\label{Fig5}}
\end{center}
\end{figure}

Figure \ref{Fig5} shows the magnetoresistance (MR = ($\rho$(B)/$\rho_0$ - 1) $\times$ 100\%) of LaRu$_2$P$_2$ at 1 K up to 22 T. At low fields, we observe the superconducting transition. We observe no sign of saturation at high magnetic fields. The value of the magnetoresistance is very large, of about 35\% at 22 T. This is of the same order as the magnetoresistance found in for example the optimally doped BaFe$_2$(As$_{1-x}$P$_x$)$_2$ and La$_{2-x}$Sr$_x$CuO$_4$\cite{hayes2016scaling,hayes2018magnetoresistance,giraldo2018scale}. We do not observe a magnetoresistance which is exactly linear, as in those cases. However, a weakly linear behavior MR$\propto$ H$^n$ with n=1.32 (purple line in the inset) provides quite a good account of the whole field dependence, except at low fields, as discussed below.

\section{Discussion}

We start by highlighting the absence of electron-electron scattering in the resistivity at zero magnetic field. The behavior at relatively high temperatures is close to linear, exactly as observed in other Fe pnictides and cuprates close to a quantum critical point. However, there is a clear saturation at low temperatures which follows very well expectations taking only into account electron-phonon (high temperatures) and electron-defect (low temperatures) scattering. There is also no sign for a T$^2$ dependence, observed in Fermi liquid systems with strong mass renormalization such as heavy fermions when these are far from a quantum critical point\cite{coleman2015introduction}. Thus, transport experiments show that LaRu$_2$P$_2$ is a simple metal, in agreement with the previously observed band-structure properties that can be explained within nearly free electron calculations\cite{moll2011quantum,razzoli2012bulk,blackburn2014fermi,ying2010isotropic,karaca2016first}.

However, the magnetoresistance deviates from the usual saturating or H$^2$ behavior found in simple metals (Figure \ref{Fig5}). To study this in more detail, let us first consider electrons on the Fermi surface of LaRu$_2$P$_2$ in two separate groups, belonging to electron-like and hole-like bands\cite{han2008generic}. The MR can then be described using using the two-band galvanomagnetic effect model\cite{noto1975simple}:
\begin{equation}
\frac{\Delta\rho}{\rho_0} = \frac{\sigma_1\sigma_2(\mu_1+\mu_2)^2H^2}{(\sigma_1+\sigma_2)^2+(\sigma_1\mu_2-\sigma_2\mu_1)^2H^2}\label{Eq1},
\end{equation}
where $\sigma$  is the conductivity, $\mu$ is the mobility and the subscripts 1 and 2 refer to the two types of carriers. Considering that electrons and holes have similar scattering rate and effective mass\cite{moll2011quantum}, the average mobility can be written as $\mu$ = e$\tau_e$/m$^*_e$ $\approx$ e$\tau_h$/m$^*_h$, and therefore $\sigma_2/\sigma_1 \approx$ n$_2$/n$_1$ = $\alpha$, $\alpha$ being the ratio between the concentration for majority carriers n$_2$ and the concentration for minority carriers n$_1$. Then, Eq.\ref{Eq1} can be simplified to

\begin{equation}
\frac{\Delta\rho}{\rho_0} = \frac{4\alpha\mu^2H^2}{(1+\alpha)^2+(1-\alpha)^2\mu^2H^2}\label{Eq2},
\end{equation}
Red line in figure \ref{Fig5} is the fit to the MR using Eq.\ref{Eq2}. We find that the fit does not reproduce well the experimental data neither at low nor high fields (see inset of figure \ref{Fig5} in logarithmic scale). 

Next we consider the effect of electron-electron correlations. The above formula has been modified by Balseiro and Falicov to account for the presence of a charge or spin density wave\cite{balseiro1985density}. These electronic orders can indeed considerably modify the magnetoresistance as reported in some pnictides\cite{ping2009magnetoresistance} and other CDW systems such as KMo$_6$O$_{17}$ and NbSe$_3$\cite{tian2002magnetoresistance,shen2003magnetoresistance}. The Fermi surface is reconstructed into small pockets within a charge or spin ordered state. Some of these small ungapped pockets might be destroyed under strong magnetic field due to Landau quantization, resulting in a more effective nesting of the Fermi surface and a large MR\cite{balseiro1985density}.  It is then assumed that the magnetic field imbalances the carrier concentration n$_2$/n$_1$ to vary linearly with the magnetic field, i.e. n$_2$/n$_1$ = $\alpha$+$\beta$H with $\alpha$ the ratio of carrier concentrations at zero magnetic field and $\beta$ a constant that quantifies the magnetic field effect on the the ratio of carrier concentrations. Using this, Eq.\ref{Eq2} is modified to

\begin{equation}
\frac{\Delta\rho}{\rho_0} = \frac{4(\alpha-\beta H)\mu^2H^2}{(1+\alpha-\beta H)^2+(1-\alpha+\beta H)^2\mu^2H^2}\label{Eq3},
\end{equation}
Blue line in figure \ref{Fig5} is the fit to the MR data using Eq.\ref{Eq3}. Even if there is no charge or spin order in LaRu$_2$P$_2$, we have tried the best fit to our data, obtaining some improvement with respect to Eq.\ref{Eq2} but it is still not satisfactory, particularly at low magnetic fields.

Having eliminated electronic correlations and charge or spin density wave orders as the origin for the large magnetoresistance, we should consider a further aspect of fermiology in the electronic transport under magnetic fields. The Fermi surface of LaRu$_2$P$_2$ has an important difference with respect to Fermi surfaces of Fe based systems. As shown in Refs.\cite{moll2011quantum,razzoli2012bulk} by DFT calculations, angular resolved photoemission and quantum oscillations, there is a highly intricate and large Fermi surface that touches the Brillouin zone border at many points. It is located around the corners of the Brillouin zone, but has branches that touch each other close to the in-plane surfaces of the Brillouin zone. It has a strongly 3D shape and the orbital character stems from Ru-4d and La electrons. Such a Fermi surface has open orbits, which are known to strongly increase the high field magnetoresistance\cite{pippard1989magnetoresistance,abrikosov2017fundamentals,kapitza1929change}. Taking open orbits into account, we can modify Eq.\ref{Eq2} by including a field independent small contribution from the open orbits $\delta\sigma$ to the total conductivity $\sigma$ and write\cite{wu2020huge}:
\begin{equation}
\frac{\Delta\rho}{\rho_0}=\!\frac{(1+\alpha)^2[(1-\delta)(1+\eta^2)+\delta(1+\eta^2)^2-1]-(1-\alpha)^2\eta^2}{(1+\alpha)^2[\delta(1+\eta^2)+1]+(1-\alpha)^2\eta^2}\label{Eq4},
\end{equation}
with $\eta$ = $\mu$ H. Note that Eq.\ref{Eq4} with $\delta$ = 0 simplifies to Eq.\ref{Eq2}. Green line in figure \ref{Fig5} is the fit to the MR using Eq.\ref{Eq4}. We find a perfect agreement with the data. From the fit to Eq.\ref{Eq4}, we obtain $\alpha$ = 23.3$\pm$0.5, $\mu$ = 0.110$\pm$0.001 m$^2$V$^{-1}$s$^{-1}$ and $\delta$ = (425$\pm$7)10$^{-4}$. In LaRu$_2$P$_2$, Hall effect measurements have shown that the carrier density is electron-type and with an estimated value of n = 3$\times$10$^{22}$ cm$^{-3}$, which is an order of magnitude larger than in electron doped pnictide superconductors\cite{ying2010isotropic}. Our fitting is consistent with this result and shows that electron carrier concentration n$_e$ is 23.3 times larger than the hole carrier concentration n$_h$. The scattering time for electron-defect scattering $\tau$ = 6.2$\times$10$^{-13}$ s can be obtained from the mobility using m$^*\approx$ m$_e$\cite{moll2011quantum,razzoli2012bulk}. This value is of the same order, as the result from the Drude model, suggesting that the scattering processes are nor fundamentally modified by the magnetic field. Instead, the finite value of $\delta$ shows an important contribution to the MR from open orbits.

It is interesting to discuss the size of the magnetoresistance in terms of scattering rates. Within a simple Drude picture, we can reasonably define a scattering rate $\hbar/\tau$. The resistivity is proportional to $\hbar/\tau$, which can be estimated through the proportionality constant of a linear temperature and field increase. Namely, $\frac{d\rho(T)}{dT} \frac{1}{k_B}\propto \hbar/\tau_{th}$ and $\frac{d\rho(B)}{dB} \frac{1}{\mu_B}\propto \hbar/\tau_{field}$ with k$_B$ is the Boltzmann constant and $\mu_B$ the Bohr magneton. It turns out that both quantities in LaRu$_2$P$_2$ provide similar values ($\frac{d\rho(T)}{dT} \approx 10 \frac{\mu\Omega cm}{meV}$ and $\frac{d\rho(B)}{dB} \approx 2 \frac{\mu\Omega cm}{meV}$). This suggests indeed that the scattering rate is high and the quasiparticle lifetime is short at high temperatures and under high magnetic fields. The situation is opposite to Refs \cite{hayes2016scaling,hayes2018magnetoresistance,giraldo2018scale,legros2019universal}. There, the magnetic field and temperature ratios are similar at temperatures where the quasiparticle lifetime and the mean free path are large. In LaRu$_2$P$_2$, these ratios equalize when the mean free paths are very short, because of strong Umklapp scattering with phonons at high temperatures and because of strong scattering due to the peculiar orbital structure at high magnetic field. Umklapp scattering is proportional to the phonon occupation, which increases linearly with temperature, and the combination of conduction paths due to intricate orbits is equivalent to introduce an effective scattering rate that increases nearly linearly with the magnetic field. While the coincidence between $\frac{d\rho(B)}{dB}$  and $\frac{d\rho(T)}{dT}$  has deep implications in the Fe based pnictides and in the cuprates and can be associated to field and temperature interwoven behaviors due to electronic correlations, spin density waves or dynamic critical fluctuations in proximity to a quantum critical point\cite{hayes2016scaling,hayes2018magnetoresistance,giraldo2018scale,legros2019universal}, in LaRu$_2$P$_2$ the coincidence seems incidental. The similarity between both numbers in LaRu$_2$P$_2$ ($\frac{d\rho(T)}{dT}$  and $\frac{d\rho(B)}{dB}$) just suggests that the electronic mean free path is very short at high magnetic fields and at high temperatures, and is probably approaching interatomic distances.

\section{Conclusions}

In conclusion, we have grown single crystals of the iron-less pnictide superconductor LaRu$_2$P$_2$ and characterized its transport and magnetic properties. The temperature dependence of the resistivity is dominated by electron-phonon scattering with no significant contribution from electron-electron interactions. We have measured a  positive non-saturating MR up to 22 T with a close to linear dependence with the magnetic field. The differences in the Fermi surface of LaRu$_2$P$_2$ , which is much more intricate and has a three-dimensional sheet with open orbits, leads to the observed large magnetoresistance. Deviations with respect to the H$^2$ behavior and the order of magnitude of the MR is comparable in LaRu$_2$P$_2$ than in Fe based superconductors. Our results suggest that its origin is not due to increased scattering due to electronic correlations, neither in form of charge or spin orders nor in electron-electron scattering, but is instead the consequence of an intricate Fermi surface topology.

\section*{Acknowledgement}

We are grateful to J.G. Analytis and I.R. Fisher for sharing with us details on the sample growth of this material. We thank P.C. Canfield for teaching us flux growth. This work was supported by the Spanish Research State Agency (FIS2017-84330-R, RYC-2014-15093, CEX2018-000805-M,  MAT2017-87134-C2-2-R), by the European Research Council PNICTEYES grant agreement 679080 and by EU program Cost CA16218 (Nanocohybri), by the Comunidad de Madrid through program NANOMAGCOST-CM (Program No. S2018/NMT-4321). We particularly acknowledge SEGAINVEX at UAM for design and construction of cryogenic equipment. We also thank R. \'Alvarez Montoya and J.M. Castilla for technical support. 


%

\end{document}